\documentclass[intlimits,twoside,a4paper]{article}

\usepackage{amsmath,amssymb}
\usepackage{graphicx}
\usepackage{wrapfig}

\usepackage[T2A]{fontenc}
\usepackage[cp1251]{inputenc}
\usepackage{cmpj2}

\issue{2012}{15}{1}{13702}

\doinumber{10.5488/CMP.15.13702}

\newcommand{\be}{\begin{equation}}
\newcommand{\ee}{\end{equation}}
\newcommand{\bea}{\begin{eqnarray}}
\newcommand{\eea}{\end{eqnarray}}

\title[Phase transitions and dynamical properties]
{Phase transitions and dynamical properties \\ of quasi-one-dimensional structures \\ with hydrogen bonds }

\author{R.Ya.~Stetsiv}

\authorcopyright{R.Ya.~Stetsiv, 2012}

\address{Institute for Condensed Matter Physics
of the National Academy of Sciences of Ukraine,\\
1 Svientsitskii Str., 79011 Lviv, Ukraine}

\date{Received November 7, 2011, in final form January 26, 2012}

\begin{document}

\maketitle

\begin{abstract}
The frequency dependence of dynamical conductivity of the quasi-one-dimensional structures with hydrogen bonds
is studied on the basis of pseudospin-electron model. It takes into account the proton-electron interaction,
external longitudinal field $h$, the tunneling hopping of protons, electron transfer and direct interaction
between protons. The dependences of the electron concentration and mean number of protons at the site on temperature
and external field are obtained.  The phase transition lines from uniform phase into charge ordered phase are
determined. The dependence of dynamical conductivity on temperature and field $h$ and its changes at the
phase transitions are obtained.

\pacs 72.60.+g, 36.40.c
\end{abstract}

\section{Introduction}

\looseness=-1The properties of molecular and crystalline structures with hydrogen bonds are mainly determined by the
character of proton redistribution on the bonds. We investigate the microscopic mechanisms of charge
transfer in such systems on the basis of the proposed pseudospin-electron model  \cite{StetS,StetDS} that takes into
account the correlation between the proton displacement and reconstruction of electron states as well as the change
of their occupancy. This interaction manifests itself as a cooperative proton-electron transfer (PET) in a
number of experimental works  \cite{Okan,Mitani,Matsush,Morimoto,Okaniwa,Takeda,Nakas,Inabe} and it follows also
from the results of quantum-chemical calculations  \cite{StasSiz,StetS,Hillenb,Scheiner}. Quantum chemical
methods allow us to examine these charge redistributions more in detail. The structural and optical studies of the
proton transfer in $N$-salicylideneaniline~\cite{Takeda,Inabe} show that photochromism and thermochromism in
these object arise from a proton transfer that is accompanied by a configurational change of electron
structure. It was shown that the behaviour of proton dynamics is quite consistent with the temperature dependence of
visible absorption spectra of this crystal. If we could construct a molecular conductor based on this type of
molecules, the charge transport might strongly be modulated by the proton motion. Photoinduced proton-coupled
electron transfer (PCET) is investigated in a number of works  \cite{Zhao,Cukier,Fang1,Fang2} as one of the
mechanisms of energy transformation in biological and chemical systems. The effect of a such proton-electron
coupling plays an important role in passing a proton through the biological membrane in photosynthesis. The
design of an electron-proton hybrid system using the elements of one-dimensional metal chains,
acceptor (or donor) molecules, and interchain H-bonds are proposed  \cite{Mitani}. A new molecular function is
expected to be produced in this system, if the motion of proton is closely correlated with the dynamics of the
1D electronic states. A similar effect is observed in the halogen (X)-bridge mixed-valence transition-metal
(M) complexes (M-X-complexes)  \cite{Okan}. The M-X-complexes [MA$_{2}$X]Y$_{2}$ (M${}={}$Pt, Pd or Ni) have
a one-dimensional (1D) chain structure and adjacent chains are connected by hydrogen bonds. Here X stands for
a bridging halogen ion (X${}={}$Cl, Br or J), A for a ligand molecule (e.g. ethylenediamine,
cyclohexenediamine), and Y for a counter anion (e.g. Y${}={}$Br$^{-}$, ClO$_{4}^{-}$). The location of the protons
on N-H-Y induce additional electron charges on the ions M and at some conditions they form a
charge-ordered state (CDW)  \cite{Okan}. It is pointed out that the electron-proton coupling is capable of controlling the
CDW state  \cite{Matsushita}.

Pseudospin-electron model was originally proposed to describe the correlated proton-electron charge transfer in
a single complex with hydrogen bond  \cite{StetS}. This model was later extended to the description of the charge transfer
in the above mentioned quasi-one-dimensional structures with hydrogen bonds. We examined uniform phases
 \cite{StetDS}. In this work we investigate the phase transition from uniform phase into charge-ordered phase in
such systems. We study thermodynamic properties and the frequency dependence of dynamical conductivity and its
changes at the phase transitions.

\section{Hamiltonian}

The Hamiltonian of quasi-one-dimensional structures which contains chains with hydrogen bonds are written down in the
form  \cite{StetDS}:
\bea
 H_{\mathrm{eff}} &=& \sum\limits_{l} \sum\limits_{i,\sigma} \left\{ (\varepsilon - \mu)n_{i\sigma}(l) +
g\left[n_{i\sigma}(l) - n_{i+1,\sigma}(l)\right]S_i^z (l)\right\}   \nonumber\\
&& + \sum\limits_{l,l'} \sum\limits_{i, j, \sigma} t_{i(l),j(l')}\left[a_{i\sigma}^{+}(l)a_{j,\sigma}(l')
 + a_{j,\sigma}^{+}(l')a_{i\sigma}(l)\right] \nonumber\\
&& + \sum\limits_{l} \sum\limits_{i} \Omega S_i^{x}(l) \nonumber\\
&& - \sum\limits_{l} \sum\limits_{i} h S_i^{z}(l) \nonumber\\
&&-\frac12\sum\limits_{l,l'} \sum\limits_{i,j} J_{i(l),j(l')}
S_i^z(l)S_j^z(l'). \label{e1}\eea
Here, the summation along the chains (indices $i$, $j$) and the summation over the chains (indices $l$, $l'$) is
performed. Pseudospin operator  $\hat S_i$ describes the proton position in double potential well on the hydrogen
bond. We suppose that the transfer along hydrogen bond is dominant: $t = t_{i(l),i+1(l)}$;
 $n_{i\sigma}$ is operator of electron concentration at $i$ lattice site,
$\sigma$ is electron spin, $\mu$ is chemical potential of electrons.

The Hamiltonian includes proton-electron interaction (parameter $g$), electron transfer (parameter $t$), energy
of proton tunneling (parameter $\Omega$), asymmetry of the local anharmonic potential (parameter $h$). The last
term describes proton-proton interaction.

Pseudospin-electron interaction leads to the effective interaction between pseudospins (between protons in our
case) and as it is shown in~\cite{StetSS,StetSS_2,StetSSS} it can cause the appearance of a modulated phase with
doubling of the initial lattice period and can lead to the corresponding charge modulation. The study of this
phenomenon is the aim of this paper. In a case of double modulation of the lattice period, the crystal is divided
into two sublattices. We introduce the following notations: $\eta_{\alpha} = \langle S_{i,\alpha}^z \rangle$,
$n_{\alpha} = \langle \sum\limits_{\sigma}n_{i,\alpha,\sigma} \rangle$, ($\alpha = 1,2$ is the sublattice index).
In the mean field approximation (MF) and after passing to $k$-representation the Hamiltonian (\ref{e1}) has a
form:
\bea
\label{e2}
&& H_{\mathrm{MF}} = H_{\mathrm{el}} + H_{\mathrm{sp}} + U,
\eea
\[
H_{\mathrm{el}} = \sum\limits_{k,\alpha,\sigma}\left[\varepsilon - \mu +
g\left(\eta_{\alpha} - \eta_{\beta} \right)\right]n_{k,\alpha,\sigma} +
\sum\limits_{k,\alpha,\sigma}t_{k,\alpha,\sigma}a_{k,\alpha,\sigma}^{+}a_{k,\beta,\sigma}\,,\qquad
\alpha \neq \beta ,
\]
\[
t_{k}^{11} = t_{k}^{22} = 0, \qquad  t_{k} \equiv  t_{k}^{12} = t_{k}^{21} =
\sum\limits_{i(l)-j(l')}t_{i(l),j(l')}^{12}\exp \left\{\ri
\vec{k}\left[\vec{R}_{i(l),1}-\vec{R}_{j(l'),2}\right]\right\},
\]
\[
H_{\mathrm{sp}} = \sum\limits_{l} \sum\limits_{i,\alpha}\left\{\Omega
S_{i,\alpha}^x(l) - \left[h + j\eta_{\beta} - g\left(n_{\alpha} - n_{\beta}
\right)\right]S_{i,\alpha}^z(l)\right\},
\]
\[
U = \frac12 N j \eta_{1}\eta_{2} - \frac{N}{2}g(n_{1} -
n_{2})(\eta_{1} - \eta_{2} ), \qquad J = \sum\limits_{l'}
\sum\limits_{j}J_{i(l),j(l')}\,.
\]

 The electronic part of Hamiltonian (\ref{e2}) is diagonalized by unitary transformation
\bea
&& a_{k,1,\sigma} = \tilde a_{k,1,\sigma} \cos
\varphi + \tilde a_{k,2,\sigma} \sin
\varphi, \nonumber\\\label{e3}
&&  a_{k,2,\sigma} = -\tilde a_{k,1,\sigma} \sin \varphi + \tilde
a_{k,2,\sigma} \cos \varphi,
\eea
\[
\cos 2\varphi = \frac{-g(\eta_{1} - \eta_{2})}{\sqrt{g^2 (\eta_{1} -
\eta_{2} )^2 + t_k^2}}\,, \qquad \sin 2\varphi = \frac{t_k}{\sqrt{g^2
(\eta_{1} - \eta_{2} )^2 + t_k^2}}\,.
\]
Thus, we obtain:
%
\bea
\label{e4}
&& H_{\mathrm{el}} =
\sum\limits_{k,\alpha,\sigma}(E_{k,\alpha} - \mu )\tilde
n_{k,\alpha,\sigma}\,,
\eea
\[
E_{k,\alpha} = \varepsilon + (-1)^\alpha \sqrt{g^2 (\eta_{1} -
\eta_{2} )^2 + t_k^2}\,.
\]

The spin part of Hamiltonian is diagonalized by unitary transformation:
\begin{eqnarray*}
 S_i^x(l) &=& \tilde S_i^x(l) \cos \psi + \tilde S_i^z(l) \sin
\psi, \\
 S_i^z(l) &=& -\tilde S_i^x(l) \sin \psi + \tilde S_i^z(l) \cos
\psi, \\
 \cos \psi_{\alpha} &=& \left[h + J \eta_{\beta} - g\left(n_{\alpha} -
n_{\beta}\right)\right]/\lambda_{\alpha}, \sin \psi_{\alpha} =
\Omega/\lambda_{\alpha}\,.
\end{eqnarray*}

In this case:
%
\bea \label{e5} H_{\mathrm{sp}} = - \sum\limits_{l} \sum\limits_{i,\alpha}
\lambda_{\alpha} \tilde S_{i,\alpha}^z(l), \eea
\[
\lambda_{\alpha} = \sqrt{\left[h + J \eta_{\beta} - g\left(n_{\alpha} - n_{\beta}\right) \right]^2 + \Omega^2}.
\]

\section{Thermodynamic properties}

Using formulae (\ref{e2})--(\ref{e5}), we can write the equations for electron concentration $n_{\alpha}$ and the
average mean of pseudospins $\eta_{\alpha}$ in sublattices:
%
\bea
\label{e6}
 n_{\alpha} &=&
\frac{1}{\frac{N}{2}}\sum\limits_{k,\alpha,\sigma}\left(\frac{1 + \cos
2 \varphi }{2}\left\{1 + \exp \left[\beta \left(E_{k,\alpha} - \mu \right) \right]\right\}^{-1}\right. \nonumber\\
&&{}\left.+\frac{1 - \cos 2 \varphi }{2}\left\{1 + \exp \left[\beta \left(E_{k,\beta} - \mu \right)
\right]\right\}^{-1}\right),
\\
%
 \label{e7}  \eta_{\alpha} &=& \frac{h + J \eta_{\beta} - g\left(n_{\alpha} - n_{\beta}\right)}{2\lambda_{\alpha}}\tanh
\left( \frac{\lambda_{\alpha}}{2kT} \right). \eea

From all the possible solutions of equations (\ref{e6})--(\ref{e7}) we choose the ones that correspond to the
minimum of grand canonical potential  $\Phi$ in regime of $\mu = {}$const
 or minimum of free energy $F = \Phi + \mu N$  in regime $n ={}$
 const.
 In the MF-approximation:
%
\bea \label{e8} \Phi &=& - 2kT \sum\limits_{k}\ln \Big(\left\{1 + \exp \left[-\beta
\left(E_{k,1} - \mu \right) \right]\right\}\left\{1 + \exp \left[-\beta \left(E_{k,2} - \mu \right) \right]\right\}\Big)  \nonumber\\
&&{}-\frac12 kTN\ln \left[4 \cosh \left( \frac{\lambda_{1}}{2kT} \right)\cosh
\left( \frac{\lambda_{2}}{2kT} \right)\right]  \nonumber\\
&&{}+ \frac12NJ\eta_{1}\eta_{2} - \frac12Ng\left(n_{1} - n_{2}\right)\left(\eta_{1} - \eta_{2}\right). \eea

From the relations (\ref{e6})--(\ref{e7}) we obtain the equations for $\delta n = n_{1} - n_{2}$ and $\delta
\eta = \eta_{1} - \eta_{2}$ , which can play a role of the order parameter for a modulated phase. Using these
equations we obtain the condition of the appearance of nonzero solutions for $\delta n$ and $\delta \eta$, and
the equation for temperature of the second order phase transition to the modulated phase.
%
\bea \label{e9} \lefteqn{ \frac{2}{\lambda^2}
\left[\frac{\Omega^2}{\lambda}\langle \sigma^z \rangle +
\frac{1}{kT}\left(h + J\eta\right)^2\left(\frac14 - \langle \sigma^z \rangle^2\right)\right]
 }\nonumber\\
 &&{}\times
\left[\frac12J -\frac{4}{N} \sum\limits_k
\frac{g^2}{t_k}\delta \eta\left(\left\{1 + \exp \left[\beta \left(\varepsilon - |t_k| - \mu
\right) \right]\right\}^{-1}-\left\{1 + \exp \left[\beta \left(\varepsilon + |t_k| - \mu \right) \right]\right\}^{-1}\right)\right] + 1 = 0\,.
\nonumber\\
\eea
Here:
\[
n = \frac{n_1 + n_2}{2}\,, \qquad \eta = \frac{\eta_1 + \eta_2}{2}\,,\qquad  \langle \sigma^z \rangle = \frac12 \tanh \left(
\frac{\lambda}{2kT} \right),\qquad
\lambda = \sqrt{(h + J \eta)^2 + \Omega^2}.
\]

\begin{figure}[ht]
\centerline{
\includegraphics[width=0.46\textwidth]{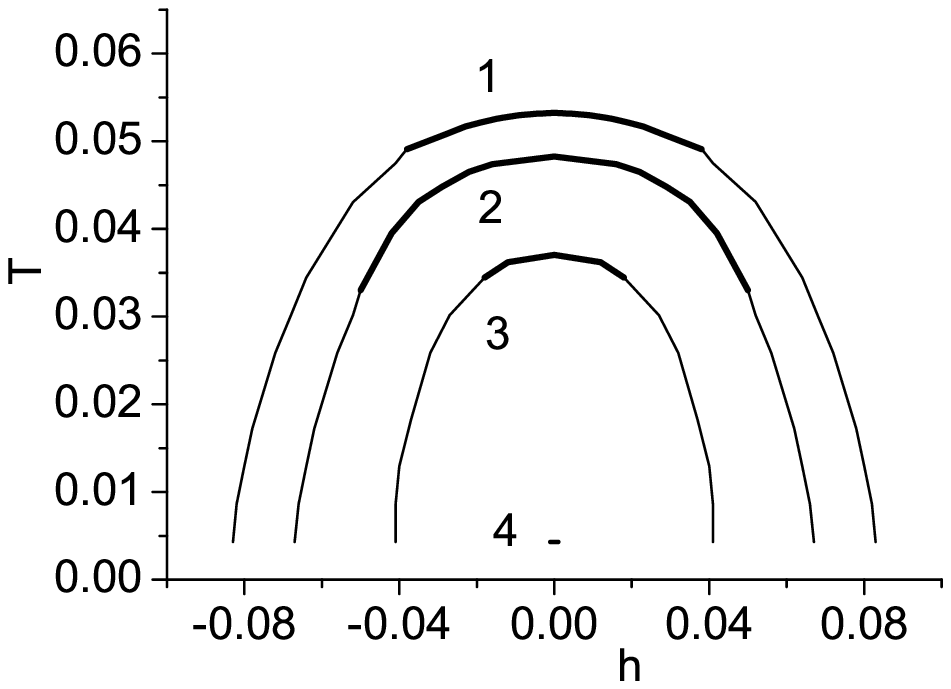} \quad
\includegraphics[width=0.46\textwidth]{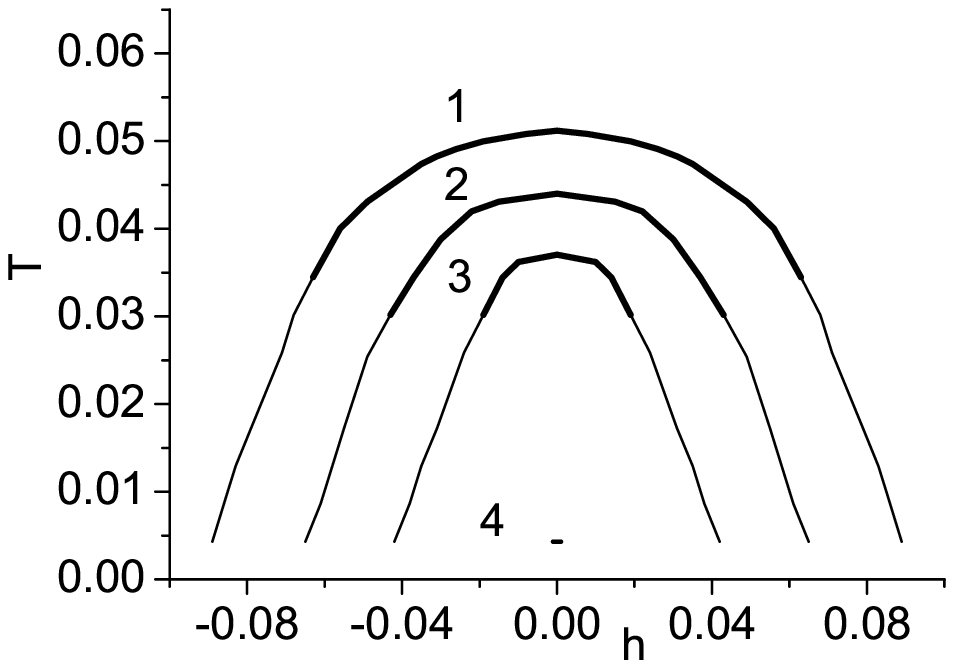}
} \centerline{ \centerline{\hspace*{1cm} (a) \hspace*{4cm} (b)}} \caption{The phase transition lines of the
first and second order from the uniform phase to the phase with double modulation: (a) for different values of
the chemical potential $\mu$: 1, 2, 3, 4~--- $\mu = 0, 0.05, 0.08, 0.12$~eV ($\Omega = 0$, $J = 0 $), (b) for
different values of parameter  $J$: 1, 2, 3, 4~--- $J = 0, 0.05, 0.1, 0.2$~eV ($\Omega = 0.05$~eV, $\mu = 0$).
Parameters $T, h$ are in eV  units.} \label{figrlabel1}
\end{figure}

\begin{figure}[ht]
\centerline{
\includegraphics[width=0.46\textwidth]{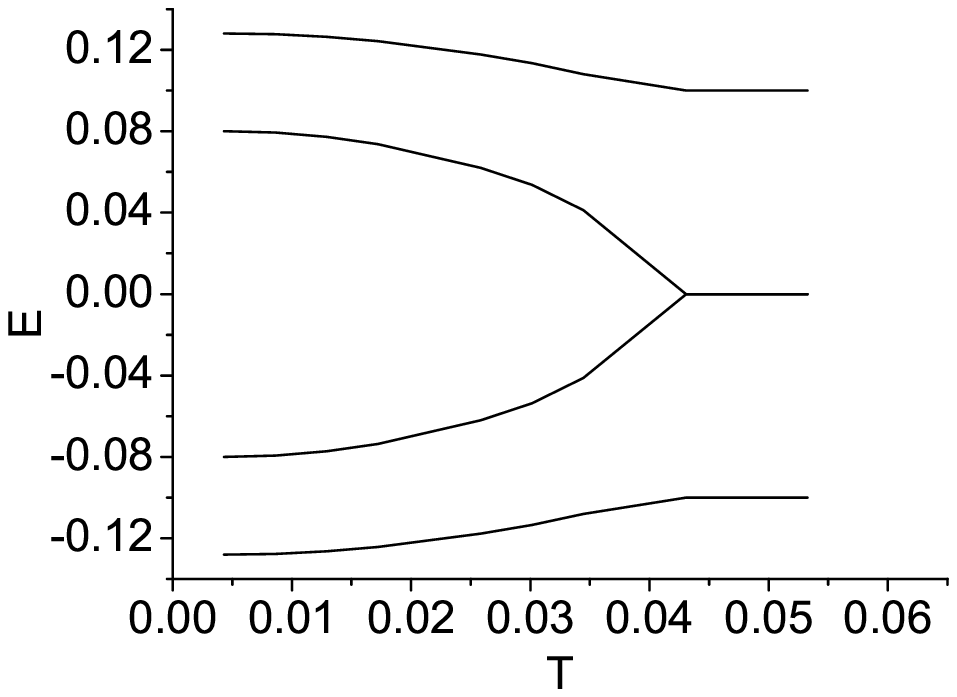} \quad
\includegraphics[width=0.46\textwidth]{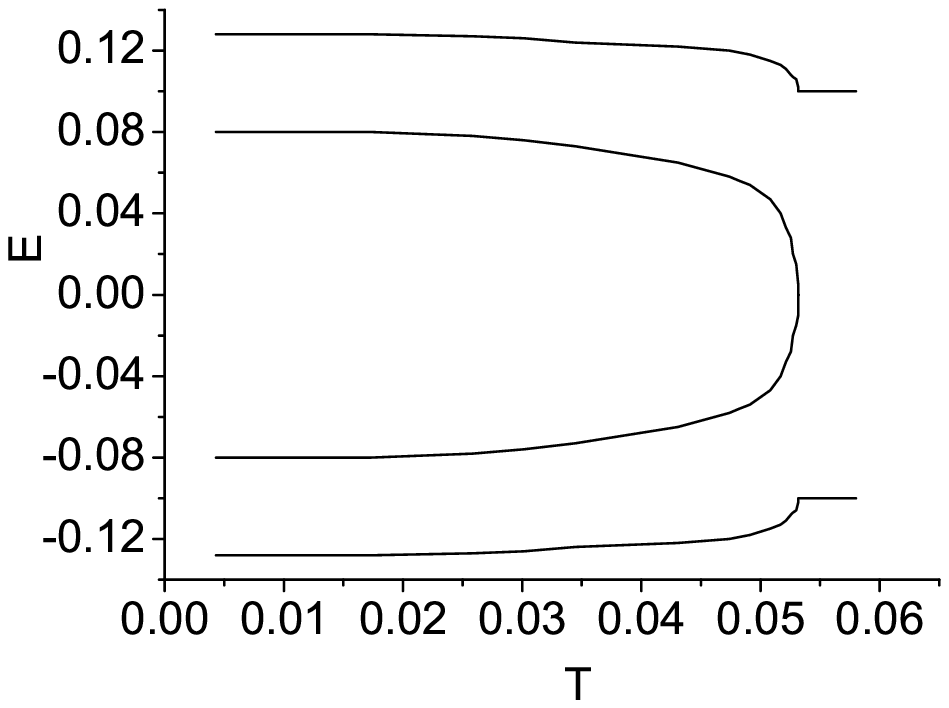}
} \centerline{ \centerline{\hspace*{1cm} (a) \hspace*{4cm} (b)}} \caption{The splitting of the electron band:
(a) along the phase transition line  (b) for  $h = 0$ in charge ordered phase, $\Omega = 0$, ($\mu = 0$, $J = 0
$). Parameters $E, T$ are in eV  units.} \label{figrlabel2}
\end{figure}

\begin{figure}[!h]
\centerline{
\includegraphics[width=0.46\textwidth]{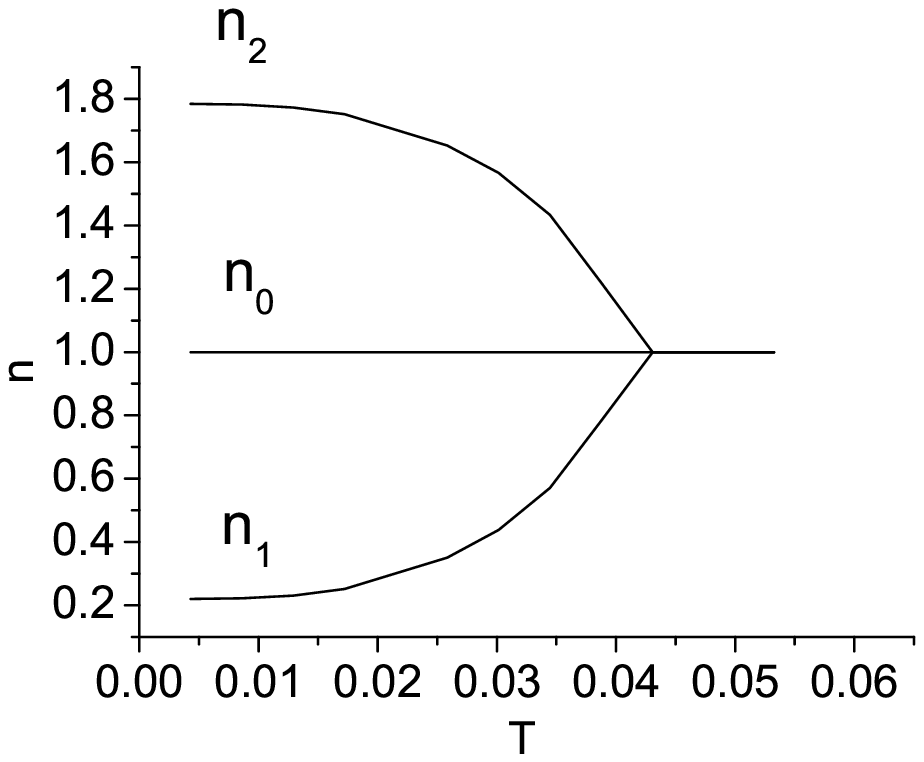} \quad
\includegraphics[width=0.46\textwidth]{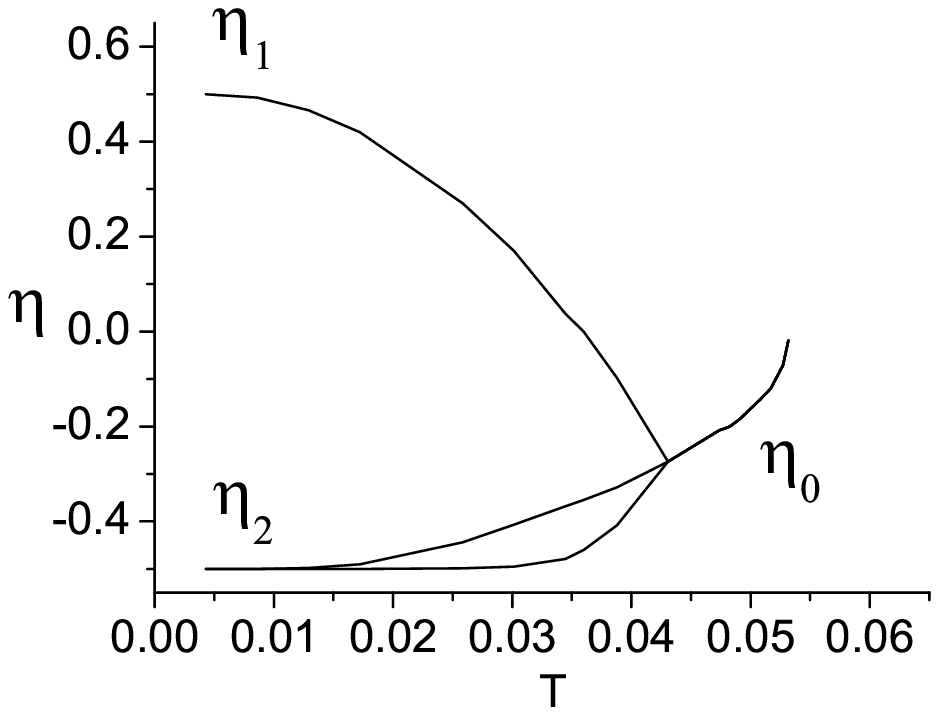}}
\caption{The temperature dependence of mean numbers of electrons of sublattice $n_{1}$, $n_{2}$, and uniform
phase $n_{0}$ and temperature dependence of mean values of pseudospins $\eta_{1}$, $\eta_{2}$, $\eta_{0}$
along the phase transition line, $\Omega = 0 $, ($\mu = 0$, $J = 0 $).} \label{figrlabel3}
\end{figure}

\begin{figure}[!h]
\centerline{
\includegraphics[width=0.46\textwidth]{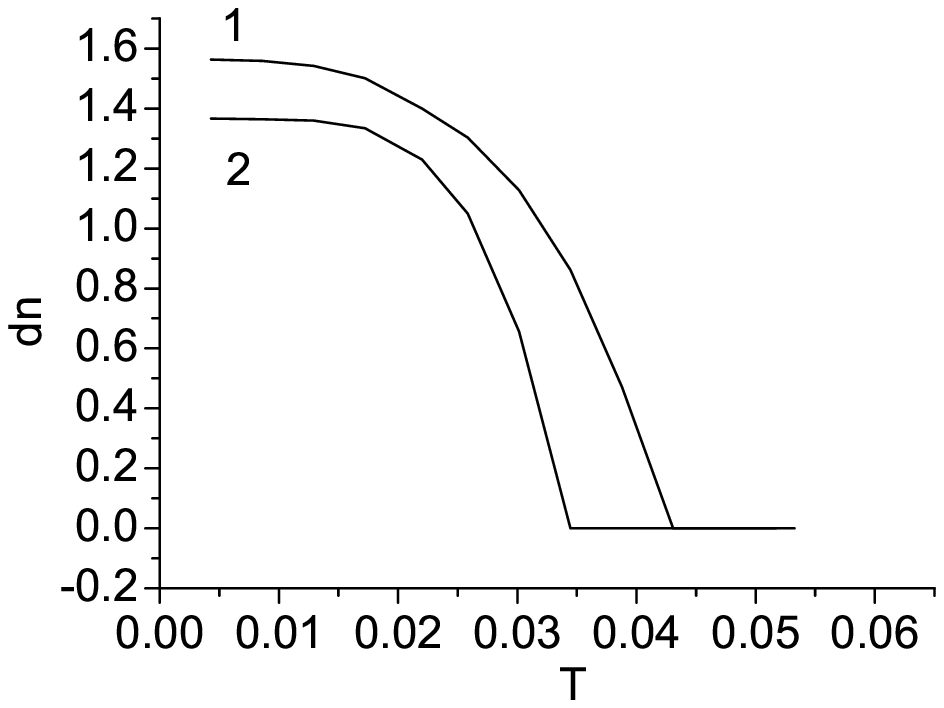} \quad
\includegraphics[width=0.46\textwidth]{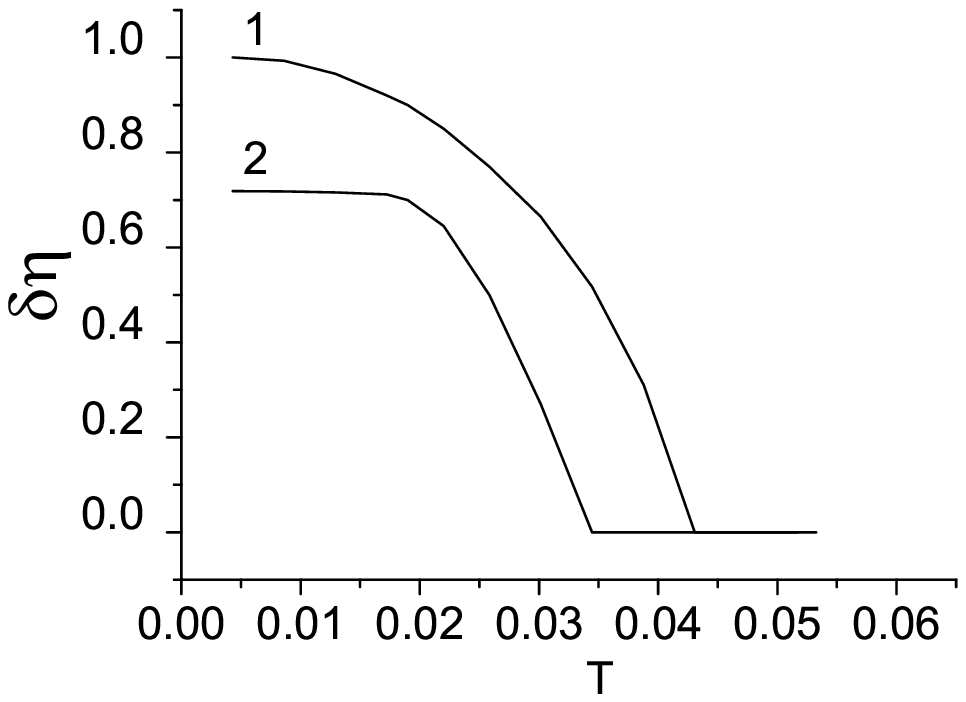}
} \caption{The temperature dependence of the values $\delta n$ and $\delta \eta$ along the phase transition
line, 1~--- $\Omega = 0$; 2~--- $\Omega = 0.05$~eV, ($\mu = 0$, $J = 0 $).} \label{figrlabel4}
\end{figure}

At certain conditions, the order of phase transition can change from the second to the first one. The phase
transition lines of the first and second order from the uniform phase to the phase with double modulation are
shown in figure~\ref{figrlabel1}: (a) for different values of the chemical potential $\mu$, (b) for different
values of parameter  $J$ and for $\Omega = 0 $ and $\Omega = 0.05$~eV cases. Transition point of the first kind
defined by numerical calculation as the point at which the requirement of minimum for thermodynamic potential
with changing the parameters of the model is transition from a homogeneous solution  $ n_ {1} = n_ {2} $,
$\eta_{1} = \eta_{2}$ to the modulated with different from zero $\delta n$ and $\delta \eta$. The lines of phase
transitions (PT) of the second order are shown bold and the lines of the first order PT are thin. The splitting of the
electron band at phase transition is shown in figure~\ref{figrlabel2}. The temperature dependences of mean numbers
of electrons of sublattice $n_{1}$, $n_{2}$, and uniform phase $n_{0}$ also, along the phase transition line are
shown in figure~\ref{figrlabel3}. The temperature dependences of mean values of pseudospins $\eta_{1}$,
$\eta_{2}$, $\eta_{0}$ are illustrated in figure~\ref{figrlabel3} as well. The temperature dependences of $\delta n(T)$
and $\delta \eta(T)$ are illustrated in figure~\ref{figrlabel4}.  These results are obtained for the following values
of parameters: $g = 0.08$~eV, $t = 0.05$~eV, $J = 0 $, $\mu = 0$, $\Omega = 0$ and $\Omega = 0.05$~eV. The
results for $J \neq 0 $, $\mu \neq0$ are presented in figure~\ref{figrlabel1}. Such a choice of the parameter
values corresponds to the ones given in~\cite{StetS}. All energy characteristics ($T, h, E, g, t, J, \mu, \Omega$) are in
eV  units.

\section{Dynamic conductivity of quasi-one-dimensional structures \\with hydrogen bonds}

 Calculation of the dynamic conductivity of the structure which possesses the chains with hydrogen bonds
 was carried out according to Kubo formula~\cite{Kubo}
%
\be \label{e10} \sigma(\omega,T) = \frac{1}{Na} \int\limits_0^{\infty} \rd t \exp [\ri(\omega + \ri\varepsilon)t]
\int\limits_0^{\beta} \rd\lambda \left\langle \hat j(t-\ri\hbar\lambda)\hat j(0)\right\rangle, \ee where $\hat j$ is the
current density operator
%
\be \label{e11} \hat j(0) = \frac{\ri}{\hbar}[\hat H,\hat d], \ee
$\hat d$ is dipole momentum operator
\[
\hat d = (-e) \sum\limits_{l} \sum\limits_i R_i(l) n_i(l) + z_{\mathrm{H}}^{\mathrm{eff}} \delta \sum\limits_{l} \sum\limits_i
S_i^z(l),
\]
that includes electronic and pseudospin (ionic) part. Here $\delta$ is the distance between equilibrium positions of
a proton on the bond, $\delta \approx 0.40$~\AA. According to quantum-chemical calculations,  the effective charge of
hydrogen $z_{\mathrm{H}}^{\mathrm{eff}}$ is equal to $z_{\mathrm{H}}^{\mathrm{eff}} \approx 0.25e$
%
\be \label{e12} j(t) = \re^{\frac{\ri}{\hbar}Ht}j(0)\re^{-\frac{\ri}{\hbar}Ht}. \ee

In the molecular field approximation, the operator of current density is split  into electronic and proton
(pseudo-spin) parts
%
\be \label{e13} \hat j = \hat j_{\mathrm{e}} + \hat j_{\mathrm{sp}}\,. \ee

The following expressions are obtained for these composites:
\bea  \hat j_{\mathrm{e}} &=& {-} \frac{2e}{\hbar}\!\sum\limits_{k,\alpha}\! \left[\frac{\partial
E_{\alpha}(k)}{\partial k_{z}}-(-1)^{\alpha}2E_{\alpha}(k)F(k)\right]a_{k,\alpha}^{+}a_{k,\alpha}\nonumber\\
&&{}+\frac{2e}{\hbar}\!\sum\limits_{k}\! F(k)\left[E_{1}(k)-E_{2}(k)\right](a_{k,2}^{+}a_{k,1}+a_{k,1}^{+}a_{k,2}), \label{e14} \\ \hat
j_{\mathrm{sp}} &=& \frac{\delta}{\hbar} \Omega z_{\mathrm{H}}^{\mathrm{eff}} \sum\limits_{l}\sum\limits_{i,\alpha}S_{i,\alpha}^{y}(l). \label{e15} \eea

 Calculation of correlation functions in the expression (\ref{e10}) with the use of the Wick's theorem yields
the following expressions for a real part of conductivity:
%
\be \label{e16} \sigma = \sigma_{\mathrm{e}} + \sigma_{\mathrm{sp}}\,, \ee
where the electronic part has a  form:
\bea \label{e17}  \sigma_{\mathrm{e}}(\omega) &=& \frac{4\pi e^2}{Na\hbar^2}\beta \sum\limits_{k,\alpha}\left[\frac{\partial
E_{\alpha}(k)}{\partial k_{z}}-(-1)^{\alpha}2E_{\alpha}(k)F(k)\right]^2 \frac{\re^{\beta\left[E_{\alpha}(k)-\mu\right]}}
{\left\{1+\re^{\beta\left[E_{\alpha}(k)-\mu\right]}\right\}^2}\delta(\omega)  \nonumber \\
&&{}+ \frac{4\pi
e^2}{Na\hbar^2}\sum\limits_{k}F(k)^2\left[E_{2}(k)-E_{1}(k)\right]\frac{\re^{\beta\left[E_{2}(k)-\mu\right]}
-\re^{\beta\left[E_{1}(k)-\mu\right]}}
{\left\{1+\re^{\beta\left[E_{1}(k)-\mu\right]}\right\}\left\{1+\re^{\beta\left[E_{2}(k)-\mu\right]}\right\}}\nonumber\\
 &&{} \times  \left[\delta\left(\omega+\frac{1}{\hbar}\left[E_{2}(k)-E_{1}(k)\right]\right)
+\delta\left(\omega-\frac{1}{\hbar}\left[E_{2}(k)-E_{1}(k)\right]\right)\right]. \eea
Here
\[
F(k) = -\frac{1}{2}\frac{g\left(\eta_{1} - \eta_{2}\right)t_{k}^{\prime}}{g^2\left(\eta_{1} - \eta_{2}\right)^2 + t_k^2}\,.
\]

For the protonic  part of conductivity we obtain:
%
\be \label{e18} \sigma_{\mathrm{sp}}(\omega) = \frac{\pi}{2a} \!\left( \frac{\delta}{2\hbar}z_{\mathrm{H}}^{\mathrm{eff}}\Omega\right)^{\!2}
\!\!\sum\limits_{\alpha}\frac{1}{\lambda_{\alpha}}
\frac{1{-}\re^{{-}\beta\lambda_{\alpha}}}{1{+}\re^{{-}\beta\lambda_{\alpha}}} \left[
\delta(\omega{-}\lambda_{\alpha}/\hbar) + \delta(\omega{+}\lambda_{\alpha}/\hbar) \right].
 \ee
\begin{figure}[!h]
\includegraphics[width=0.49\textwidth]{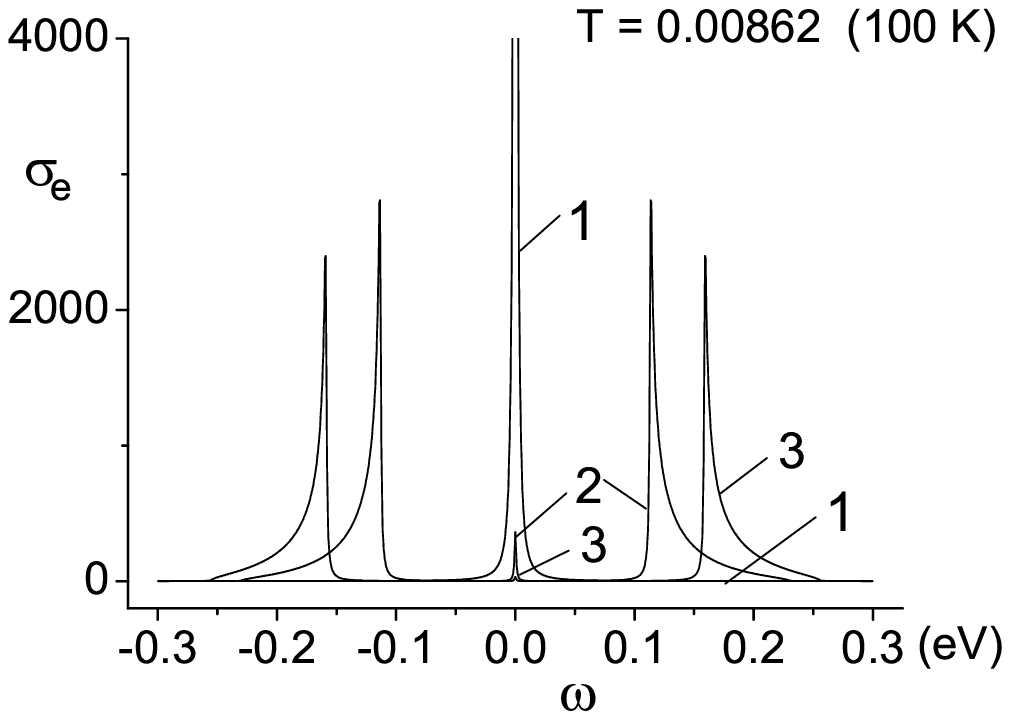}%
\hfill%
\includegraphics[width=0.49\textwidth]{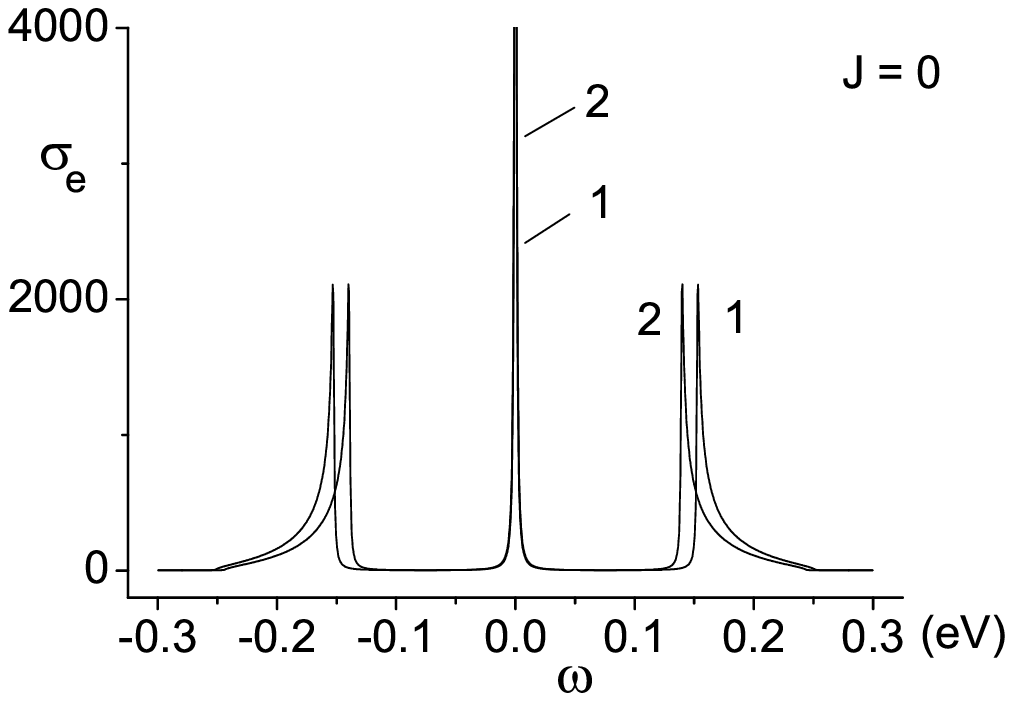}%
\\%
\includegraphics[width=0.49\textwidth]{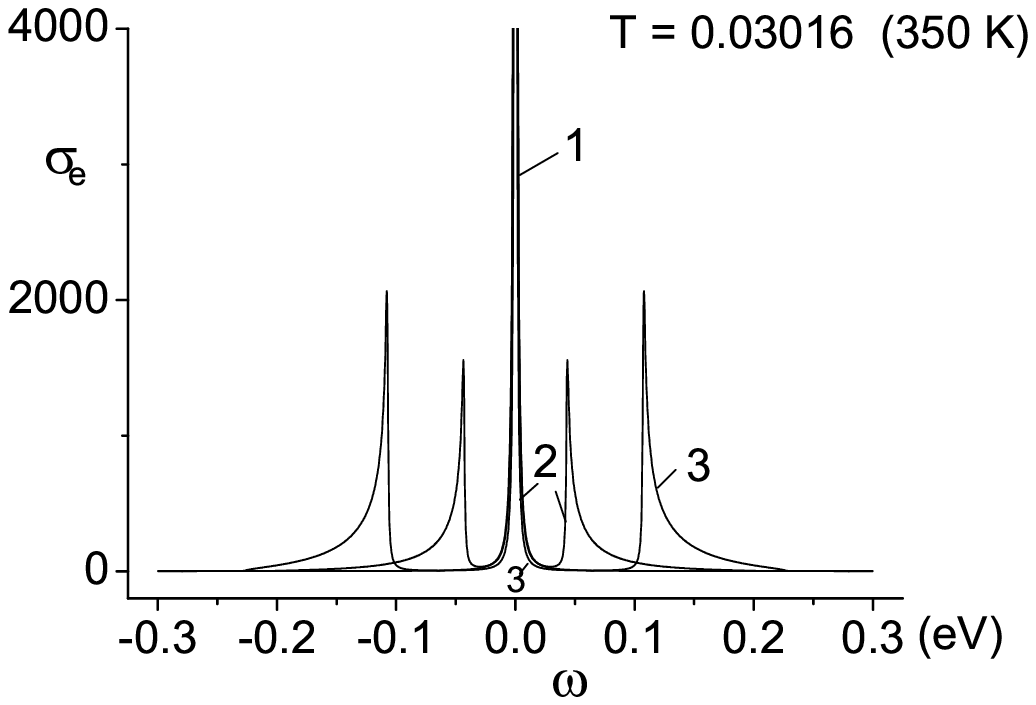}%
\hfill%
\includegraphics[width=0.49\textwidth]{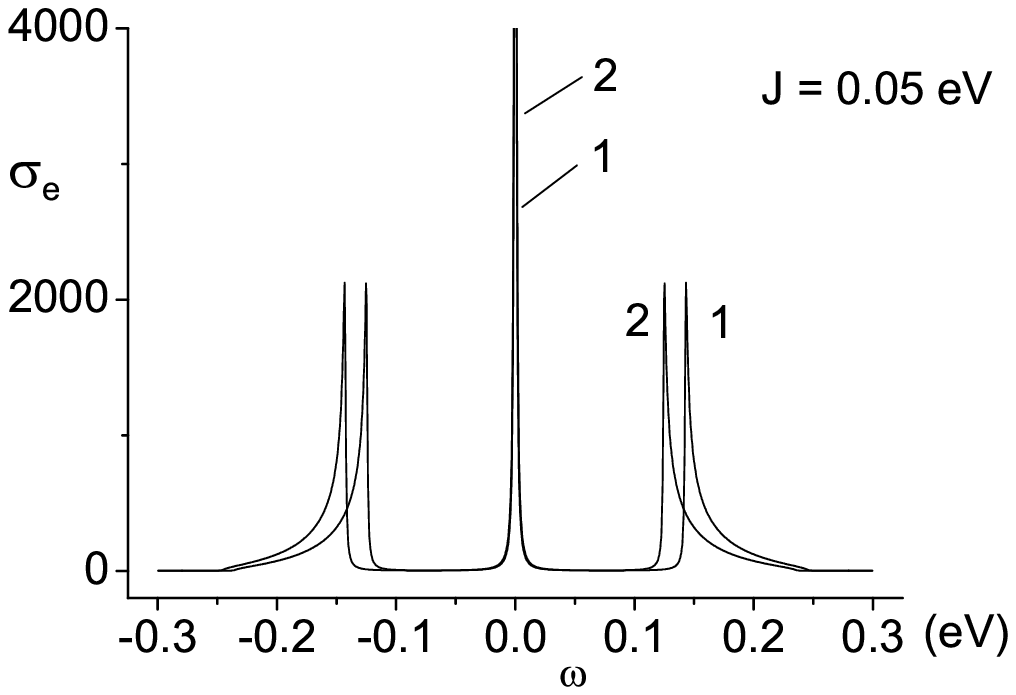}%
\\%
\includegraphics[width=0.49\textwidth]{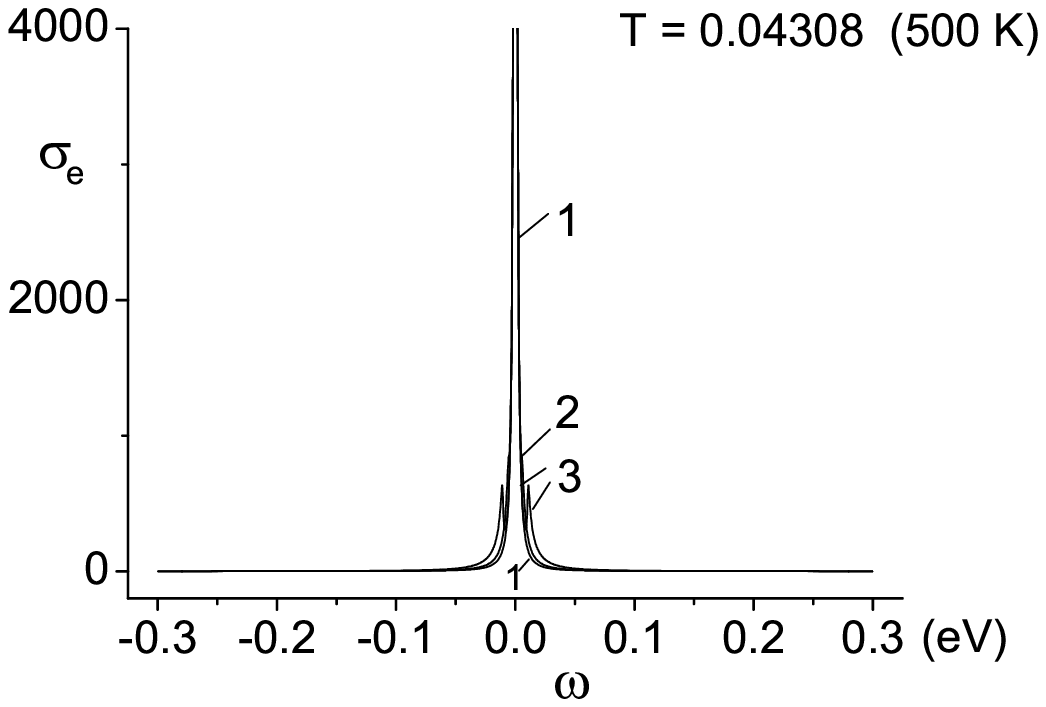}%
\hfill%
\includegraphics[width=0.49\textwidth]{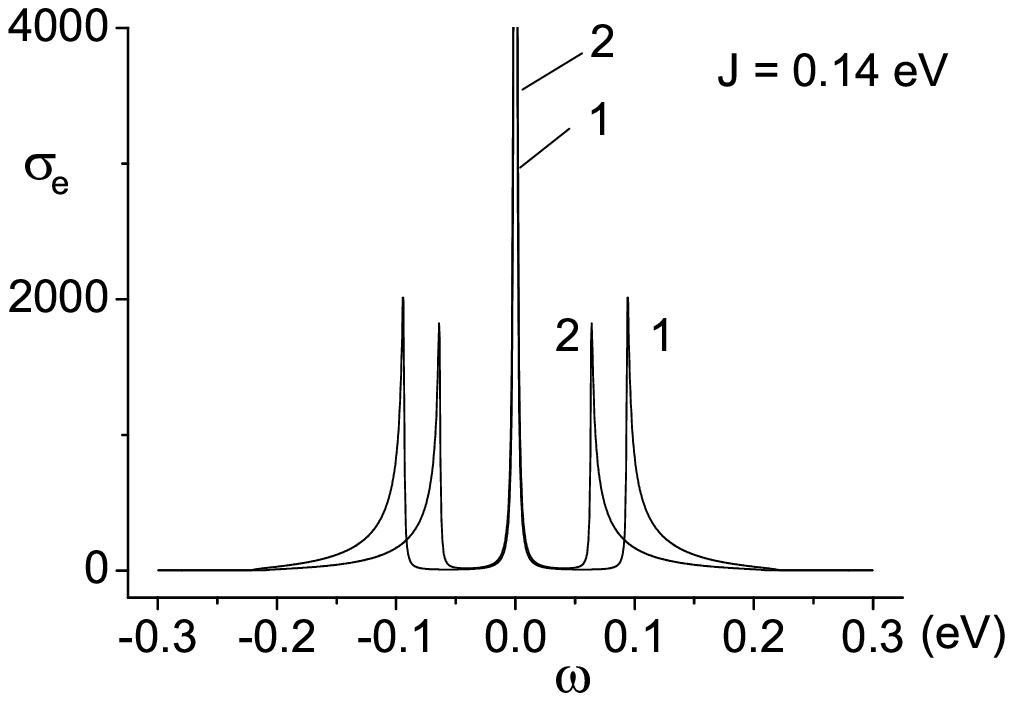}%
\\%
\parbox[t]{0.48\textwidth}{%
\caption{%
Frequency dependence of the
electronic part of conductance: 1~--- before phase transition (uniform phase),  2, 3~--- after phase transition
(modulated structure) along the phase transition line; 2~--- $\Omega = 0.05$~eV, 3~--- $\Omega = 0$; $\mu = 0$,
$J = 0$.} \label{figrlabel5}%
}%
\hfill%
\parbox[t]{0.48\textwidth}{%
\caption{Frequency dependence of the
electronic part of conductance with different values of proton-proton interaction (parameter~$J$), $h = 0$, $\mu
= 0$, $T = 0.03016$ (350~K): $J = 0; 0.05; 0.14$; 1~--- $\Omega = 0$; 2~--- $\Omega = 0.05$~eV.}
\label{figrlabel6}%
\vspace{10mm}
}%
\end{figure}

\begin{figure}[!ht]
\centerline{\includegraphics[width=0.48\textwidth]{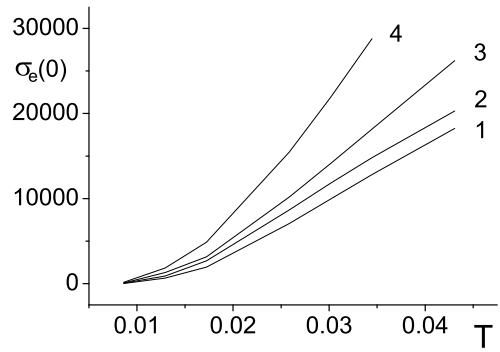}
} \caption{Temperature dependence of the electronic part of conductance of the modulated structure
 $\sigma_{\mathrm{e}}(0)$; $\mu = 0$; $h = 0$; 1, 2~--- $J = 0$; 3~--- $J = 0.05$; 4~--- $J = 0.12$; 1~--- $\Omega = 0$;
  2, 3, 4~--- $\Omega = 0.05$~eV.} \label{figrlabel8}
\end{figure}

Frequency dependence of the electronic part  of the dynamical conductivity along the phase transition line is shown
in figure~\ref{figrlabel5}, curve 1~--- before phase transition (uniform phase), curve 2, 3~--- after
 \begin{wrapfigure}{i}{0.55\textwidth}
\includegraphics[width=0.55\columnwidth]{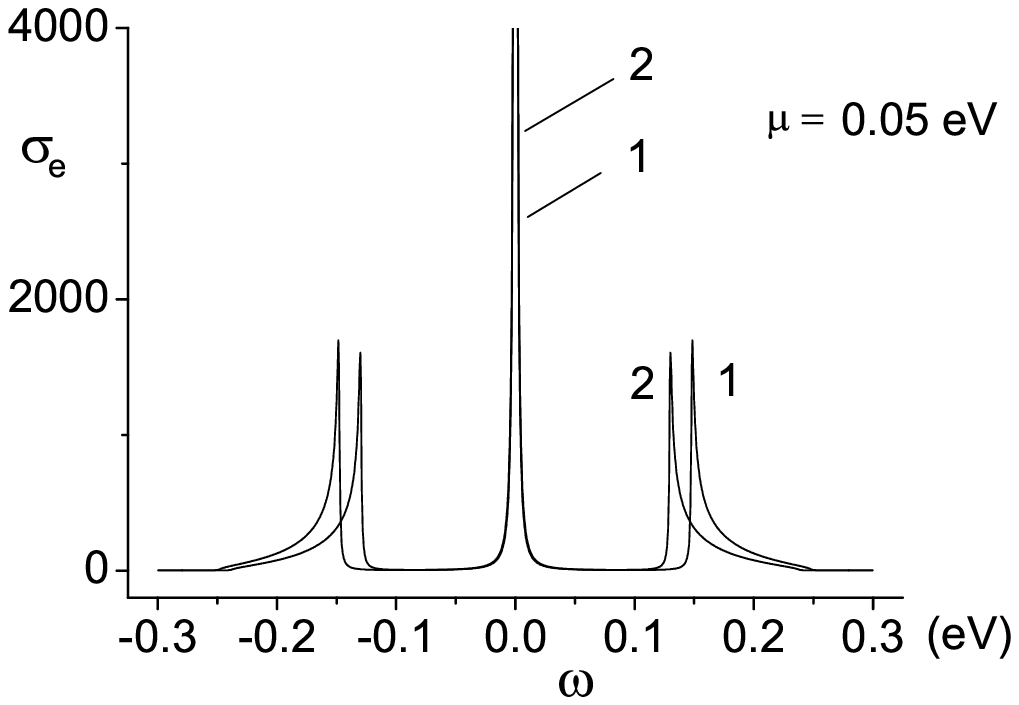}
\includegraphics[width=0.55\columnwidth]{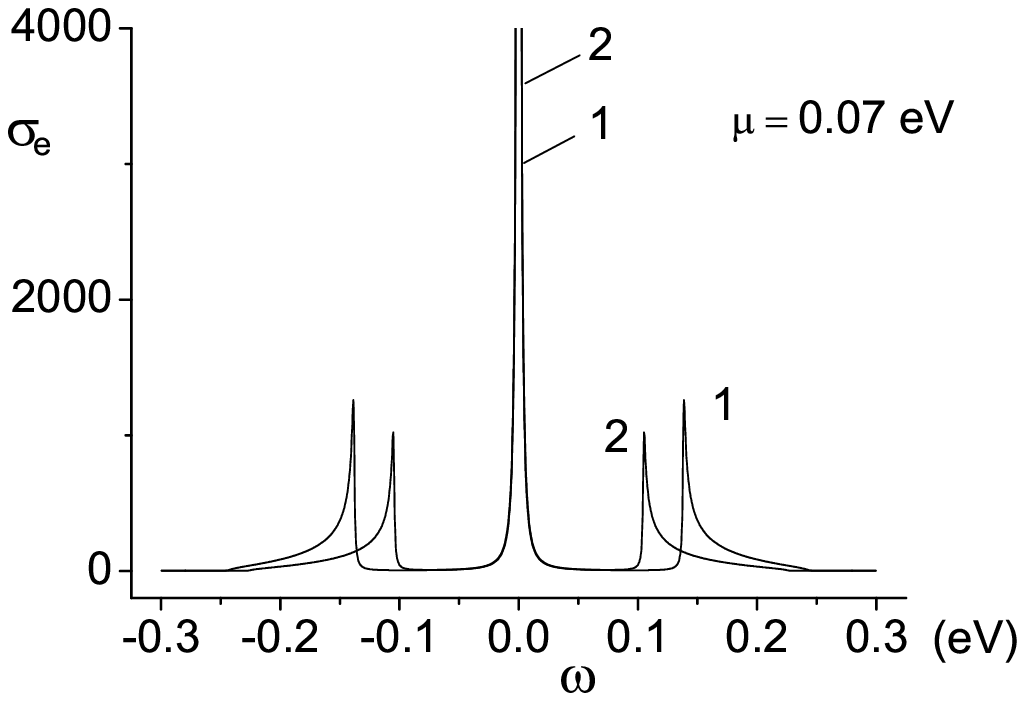}
\includegraphics[width=0.55\columnwidth]{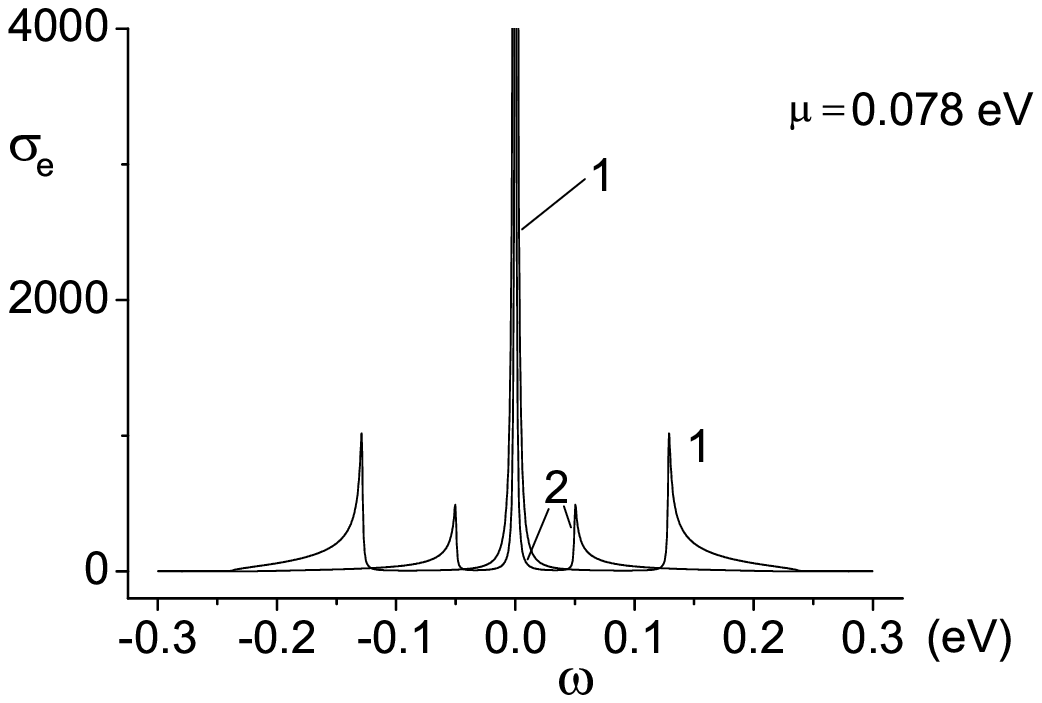}
\caption{Frequency dependence of the
electronic part of conductance with different values of chemical potential $\mu$; $h = 0$, $J = 0$,
$T = 0.03016$~(350~K): $\mu = 0.05; 0.07; 0.078$; 1~--- $\Omega = 0$; 2~--- $\Omega = 0.05$~eV.} \label{figrlabel7}
\end{wrapfigure}
phase
 transition (modulated structure).
At the phase transition from uniform to modulated structure, the conductivity $\sigma_{\mathrm{e}}(0)$, when $\omega = 0$,  is abruptly reduced by two to three orders of magnitude at low temperatures and with increasing
temperature  the value of the jump decreases.
Electronic conductivity has one peak (at $\omega = 0$) in uniform
phase, (one electronic band is present). We observed the splitting of the electron band in a modulated phase, and
 electronic conductivity has a broad maximum in the frequency region $\omega = \frac{1}{\hbar}[E_{2}(k)-E_{1}(k)]$ as
 well as a peak in $\omega = 0$. This broad maximum is placed in the lower frequency regions for structures
with the high proton tunneling frequency and stronger direct interaction between protons. The static
conductivity $\sigma_{\mathrm{e}}(0)$ in a modulated phase increases with temperature. The dynamical conductivity
$\sigma_{\mathrm{e}}(\omega)$ decreases with increasing temperature and its maximum shifts to a lower frequency region and
vanishes at the critical temperature when there is no modulation. In a homogeneous phase, only one peak remains
at $\omega = 0$. The change of the frequency dependence of  the dynamical conductivity with the parameters $J$
and $\mu$ is shown in figure~\ref{figrlabel6} and  figure~\ref{figrlabel7}, (here we consider the case $h = 0$ ).
Maximum of  the dynamical conductivity decreases and shifts to lower frequency region with an increase of $J$ and
$\mu$. Critical values of these parameters  exist (see figure~\ref{figrlabel1}) when the modulated phase
vanishes and there remains only a peak at $\omega = 0$. The temperature dependence of electronic conductivity
$\sigma_{\mathrm{e}}(0)$ is shown in figure~\ref{figrlabel8}. This part of the conductivity is higher for systems with larger
proton tunneling frequency  and it increases with an increase of the parameters $J$. Conductivity value is presented in
relative units.

\begin{figure}[!h]
\centerline{
\includegraphics[width=0.48\textwidth]{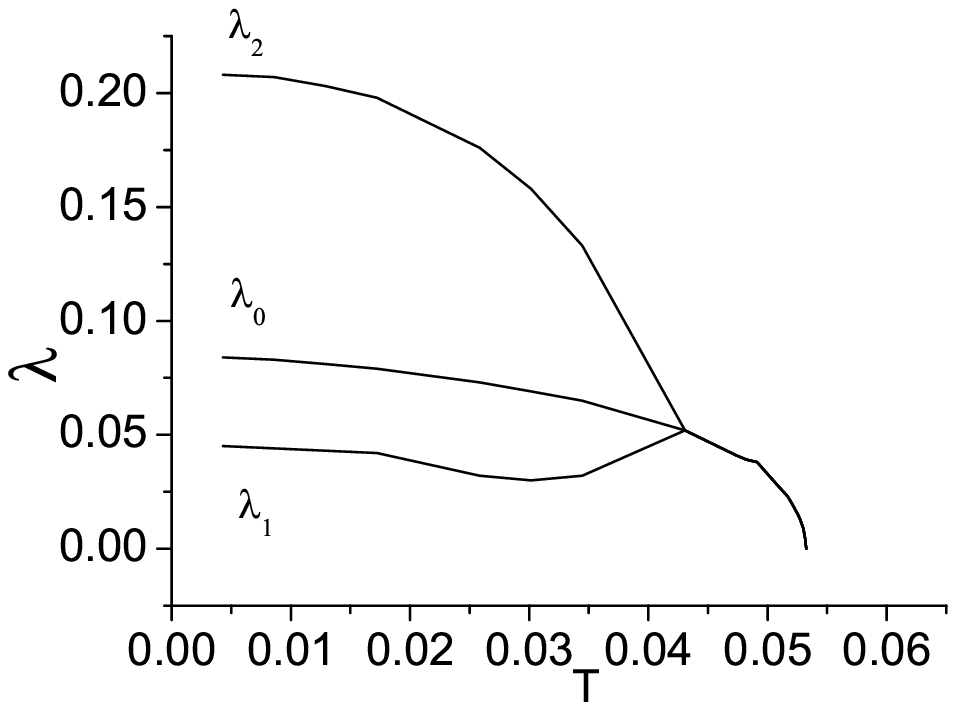}\quad
\includegraphics[width=0.48\textwidth]{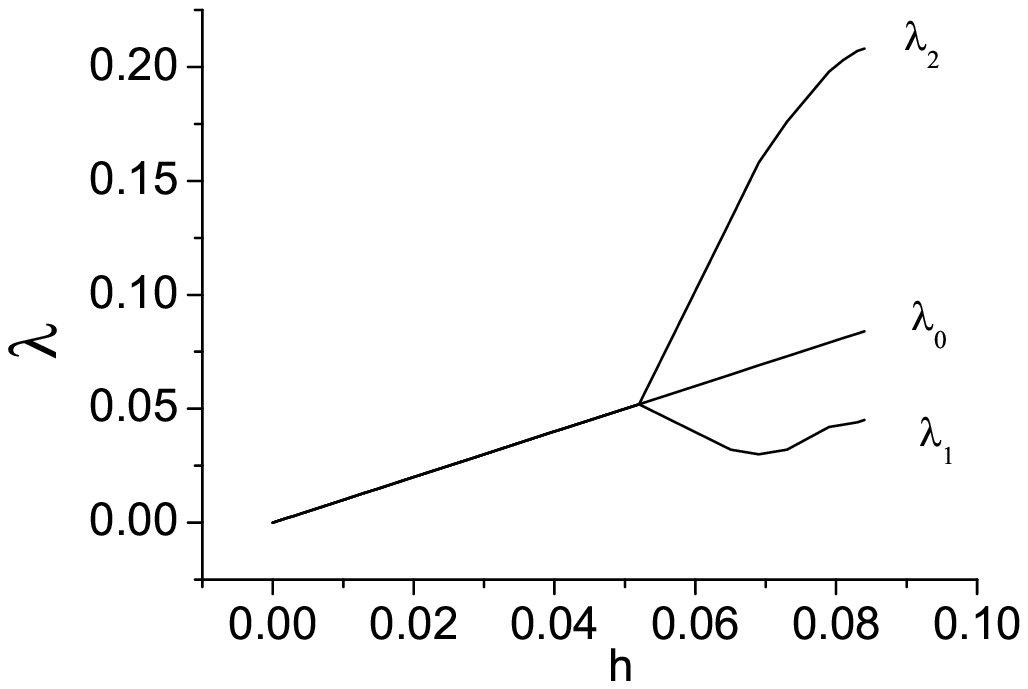}}
\centerline{(a)} \centerline{
\includegraphics[width=0.48\textwidth]{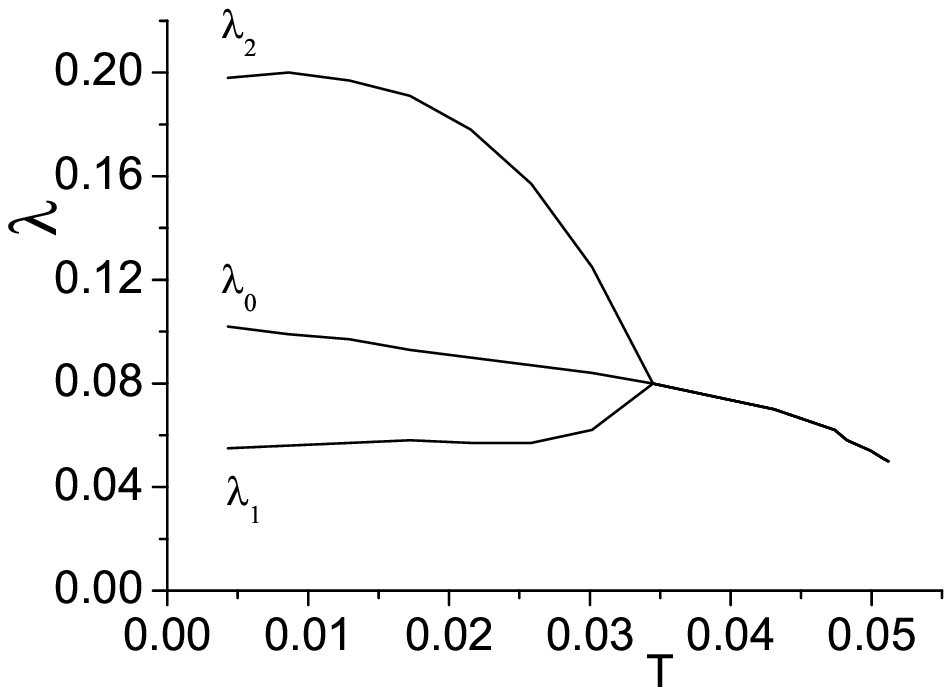}\quad
\includegraphics[width=0.48\textwidth]{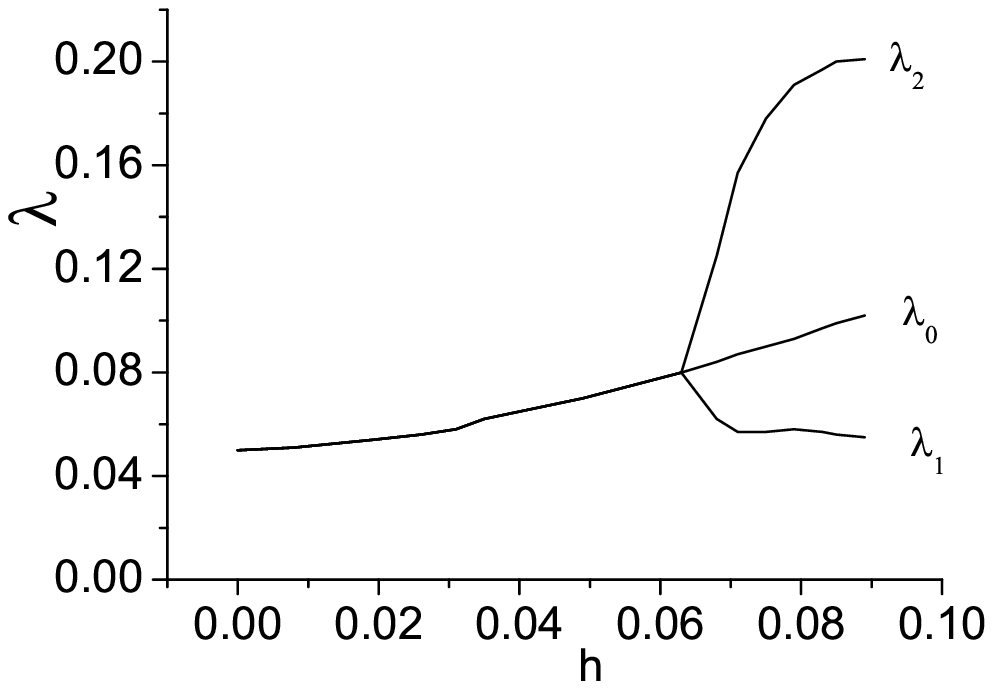}}
\centerline{(b)} \caption{The dependence of the peak-frequencies of the proton dynamical conductivity on
temperature and longitudinal field along the phase transition line;  $\lambda_{i} = \hbar \omega_{i}$; $\mu =
0$; $J = 0$; (a)~--- $\Omega = 0$; (b)~--- $\Omega = 0.05$~eV. } \label{figrlabel9}
\end{figure}

Proton dynamic conductivity has peaks at frequencies $\omega_{i} = \frac{\lambda_{i}}{\hbar}$ corresponding to
protons energies $\lambda_{i}$  on hydrogen bonds. One peak ($\lambda_{0}$) exists in case of homogeneous phase
and two peaks ($\lambda_{1}$, $\lambda_{2}$) are present for the case of a modulated structure. The dependence of the
energy $\lambda_{i}$ on temperature and asymmetry field $h$ along the phase transition line  is shown in
figure~\ref{figrlabel9}.
\newpage

\newpage

\section{Conclusions}

The possibility of the first or the second order transitions from uniform phase into phase with doubled lattice
period in the quasi-one-dimensional structures with hydrogen bonds is studied in the framework of the proposed
pseudospin-electron model. It was shown that pseudospin-electron (proton-electron) interaction may cause the
appearance of charge ordered phase in the structures with hydrogen bonds.  The electron spectrum is calculated.
The dependences of the splitting of the electron spectrum  on temperature and asymmetry field are investigated.
The dependences of the electron concentration and mean number of protons at the site on temperature and asymmetry
field were obtained. It was shown that abrupt changes of these characteristics at the first-order transitions are
smaller for the structures with high proton tunneling frequency and stronger direct interaction between protons.
The phase transition lines from uniform phase into charge ordered phase are determined. The dependences of the
dynamical conductivity on temperature and external field and its changes at the phase transitions are obtained.
At the phase transition from uniform to modulated structure the static conductivity $\sigma_{\mathrm{e}}(0)$ is abruptly
reduced by two to three orders of magnitude at low temperatures and with increasing temperature  the value of
the jump decreases. Electronic conductivity has one peak at $\omega = 0$ in a uniform phase. In modulated phase,
the dynamical electronic conductivity has a broad maximum as
 well as a peak at $\omega = 0$. This broad maximum is placed at  lower frequencies for the structures
with high proton tunneling frequency and stronger direct interaction between protons. It was shown that
the frequency dependence of the proton dynamical conductivity has one peak in a uniform phase and two peaks in the
charge modulated phase. The model can be applied to a description of quasi-one-dimensional structures, the so-called
halogen-bridge mixed-valence transition-metal complexes  \cite{Okan} in which there are charge modulated states.


\newpage

\ukrainianpart

\title{Фазовi переходи i динамiчнi властивостi квазiодновимiрних
структур з водневими зв’язками}
\author{Р.Я. Стецiв}
 \address{Інститут фізики конденсованих систем НАН України,
 вул. Свєнціцького, 1, 79011 Львів}

\makeukrtitle

\begin{abstract}
\tolerance=3000%
На основi псевдоспiн-електронної моделi дослiджено частотну залежнiсть
динамiчної провiдностi квазiодновимiрних систем з водневими зв’язками. В
моделi враховано протон-електронну взаємодiю, зовнiшнє поздовжнє поле $h$,
тунелювання протонiв, електронне перенесення i пряму протон-протонну взаємодiю.
Отримано залежнiсть електронної концентрацiї i середньої заселеностi протонних
позицiй вiд температури i поля $h$. Отримано лiнiю фазових переходiв з однорiдної
фази до фази з модуляцiєю заряду. Дослiджено залежнiсть динамiчної провiдностi вiд поля $h$ i температури та її змiни при фазових переходах.

\keywords псевдоспiн-електронна модель, протон-електронна взаємодiя,
водневий зв’язок, провiднiсть

\end{abstract}


\begin{thebibliography}{30}

\bibitem{StetS} Stasyuk I.V., Stetsiv R.Ya., Sizonenko Yu.V., Condens. Matter Phys., 2002, \textbf{5}, 685.

\bibitem{StetDS} Stasyuk I.V., Stetsiv R.Ya., Yurechko R.Ya., J. Phys. Stud., 2005,
\textbf{9}, 349.

\bibitem{Okan} Okaniwa K., Okamoto H., Mitani T., Toriumi K., Yamashita~M., J. Phys. Soc. Jpn., 1991, \textbf{60},  997; \\ \doi{10.1143/JPSJ.60.997}.

\bibitem{Mitani} Mitani T., Kitagawa H., Okamoto H., Nakasuji K., Toyota J.,
Yamashita M.,  Mol. Cryst. Liq. Cryst., 1992, \textbf{216}, 73; \doi{10.1080/10587259208028752}.


\bibitem{Matsush}
Matsushita~N., Toriumi~K., Kojima~N., Mol. Cryst. Liq. Cryst., 1992, \textbf{216}, 201; \doi{10.1080/10587259208028773}.

\bibitem{Morimoto}
 Morimoto~Y.,   Tokura~Y.,   Oohashi~T.,   Kojima~T.,   Itsubo~A.,  Mol. Cryst. Liq. Cryst., 1992, \textbf{216}, 223; \\ \doi{10.1080/10587259208028777}.

\bibitem{Okaniwa}
 Okaniwa~K.,   Okamoto~H.,   Mitani~T.,   Inabe~T.,   Tojoda~J.,   Morita~Y.,   Nakasyji~K., Yamamoto H., Deno T., Honma S.,  Mol. Cryst. Liq. Cryst., 1992,
\textbf{216}, 241; \doi{10.1080/10587259208028780}.

\bibitem{Takeda} Takeda S., Chihara H., Inabe T., Mitani T., Maruyama Y.,  Mol. Cryst. Liq.
Cryst., 1992, \textbf{216}, 235; \\ \doi{10.1080/10587259208028779}.

\bibitem{Nakas} Nakasuji K., Sugiura K., Toyoda J., Morita Y., Okamoto H.,
Okaniwa K., Mitani T.,  Mol. Cryst. Liq. Cryst., 1992, \textbf{216}, 213; \doi{10.1080/10587259208028775}.

\bibitem{Inabe}
Inabe T., Okaniwa K., Okamoto H., Mitani T., Maruyama Y.,
Takeda S.,  Mol. Cryst. Liq. Cryst., 1992, \textbf{216}, 229; \doi{10.1080/10587259208028778}.

\bibitem{StasSiz}
Stasyuk I.V., Sizonenko Yu.V., Stetsiv R.Ya.,  J. Phys. Stud., 1998,
 \textbf{2}, 463.

\bibitem{Hillenb}
Hillenbrand E.A., Scheiner S., J. Am. Chem. Soc., 1984, \textbf{106}, 6266; \doi{10.1021/ja00333a027}.

\bibitem{Scheiner}
Scheiner S.  --- In: Proton Transfer in Hydrogen-Bonded Systems. Edited by T.~Bountis. Plenum Press, New York, 1992, p.~29--50.

\bibitem{Zhao}
Zhao X.G., Cukier R.I., J. Phys. Chem., 1995, \textbf{99}, 945; \doi{10.1021/j100003a017}.

\bibitem{Cukier}
Cukier R.I., J. Phys. Chem., 1996, \textbf{100}, 15428; \doi{10.1021/jp961025g}.

\bibitem{Fang1}
Fang J.-Y., Hammes-Schiffer S., J. Chem.  Phys., 1997, \textbf{106}, 8442; \doi{10.1063/1.473903}.

\bibitem{Fang2}
Fang J.-Y., Hammes-Schiffer S., J. Chem.  Phys., 1997, \textbf{107}, 5727; \doi{10.1063/1.474333}.

\bibitem{Matsushita}
Matsushita E., Phys. Rev.~B, 1995, \textbf{51}, 17332; \doi{10.1103/PhysRevB.51.17332}.

\bibitem{StetSS}
Stasyuk I.V., Shvaika A.M., Tabunshchyk K.V., Condens. Matter Phys., 1999,
\textbf{2}, 109.
%
\bibitem{StetSS_2}
Stasyuk I.V., Shvaika A.M., Tabunshchyk K.V., Ukr. J. Phys., 2000, \textbf{45}, 520.

\bibitem{StetSSS}
Stasyuk I.V., Mysakovych T.S., J. Phys. Stud., 2001, \textbf{5}, 268.

\bibitem{Kubo}
Kubo R., J. Phys. Soc. Jpn., 1957, \textbf{12}, 570; \doi{10.1143/JPSJ.12.570}.

\end{thebibliography}
\end{document}